# Storing Energy in Biodegradable Electrochemical Supercapacitors

Guilherme Colherinhas[1], Thaciana Fileti[2] and Eudes Fileti[2]

1) Departamento de Física, CEPAE, Universidade Federal de Goiás, 74690-900, Goiânia, GO, Brazil.

2) Instituto de Ciência e Tecnologia, Universidade Federal de São Paulo, 12247-014, São José dos Campos, SP, Brazil.

* Corresponding author. E-mail: fileti@unifesp.br; fileti@gmail.com; Tel: +55 12 3309-9573;

**TOC Graph**

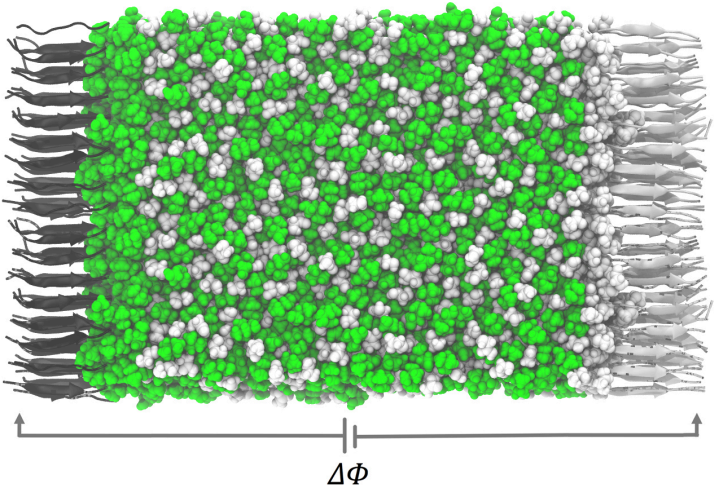

**ABSTRACT**

The development of green and biodegradable electrical components is one of the main fronts of research to overcome the growing ecological problem related to the issue of electronic waste. At the same time, such devices are highly desirable in biomedical applications such as integrated bio-electronics, for which biocompatibility is also required. Supercapacitors for storage of electrochemical energy, designed only with bio-degradable organic matter would contemplate both aspects, that is, they would be ecologically harmless after their service lifetime and would be an important component for applications in biomedical engineering. By means of atomistic simulations of molecular dynamics we propose a supercapacitor whose electrodes are formed exclusively by self-organizing peptides and whose electrolyte is a green amino acid ionic liquid. Our results indicate that this supercapacitor has a high potential for energy storage with superior performance than conventional supercapacitors. In particular its capacity to store energy was estimated to be almost 20 times greater than an analogue one of planar metallic electrodes.

The technological state in which we are immersed has increasingly required the synergistic use of environmentally friendly materials and clean energy sources. This is the case, for example, of biotechnological areas, which require materials that are both ecological and biocompatible.[1-6] Another area for which the use of environmentally and/or biocompatible materials is highly desirable is the area of sustainability and renewable energies.[7-10] In particular, for the specific case of energy storage, the biodegradability of energy storage systems after their service lifetime would drastically reduce the risks related to electronic waste.[11-13] In addition, the development of energy storage devices for bio-integrated electronics is necessarily based on biodegradable and non-toxic materials.[5-6, 14-15] In this direction, research involving natural biopolymers and small biological molecules has yielded formidable results for the efficient storage of electrochemical energy with direct applications in biomedical electronics.[7] In parallel, it has recently been shown that peptide nanostructures exhibit semiconductor characteristics that enable its use in the development of biocompatible devices for energy storage.[5] Advances in this field have already been achieved.[15-16] For example, it was observed that the efficiency of supercapacitors with carbon electrodes modified by peptide nanotubes was increased about 50-fold over the pure carbon supercapacitor. This was attributed to the expansion of the specific area of the electrode, where the hydrophilic channels of the nanotubes allow greater contact with the electrolyte.[15-16]

Recently, Hamley and collaborators have synthesized and demonstrated the high stability of bolaamphiphilic polypeptide nanosheets.[17-18] Bolaamphiphiles, unlike typical amphiphiles that have a polar head and a hydrophobic tail, present both hydrophilic ends separated by a hydrophobic core. This gives these polypeptides very interesting properties from both structural and energetic points of view. Examples of this type of system were obtained by the self-organization of the bolaamphiphilic polypeptides RFL$_4$FR and EFL$_4$FE, which are constituted by a hydrophobic core formed by a leucine tetrapeptide (L$_4$).[17-19] In this sequences the FR or FE terminations were important role; the phenylalanine residue (F) is inserted to favor the π-interactions between the lateral chains while the polar residues, arginine (R) and L-glutamic acid (E), were introduced to promote the interaction of the polypeptide sequence with the medium. The resulting nanosheets were highly ordered in the monolayer core, which gives them greater stiffness than those observed in conventional amphiphilic based nanostructures.[17-18] In addition to the remarkable structural stability, these bolaamphiphilic-based nanosheets are capable of stabilizing a high surface density of charges in contact with the medium.[17-19] In alkaline solution the termination of each polar group is -COO and -NH$_3$ so that the sequences EFL$_4$FE and RFL$_4$FR assume a net charge of -2$e$ and +2$e$, respectively (see Fig 1).

**Figure 1**: At right, molecular representation of the bolaamphiphilic polypeptides and their net

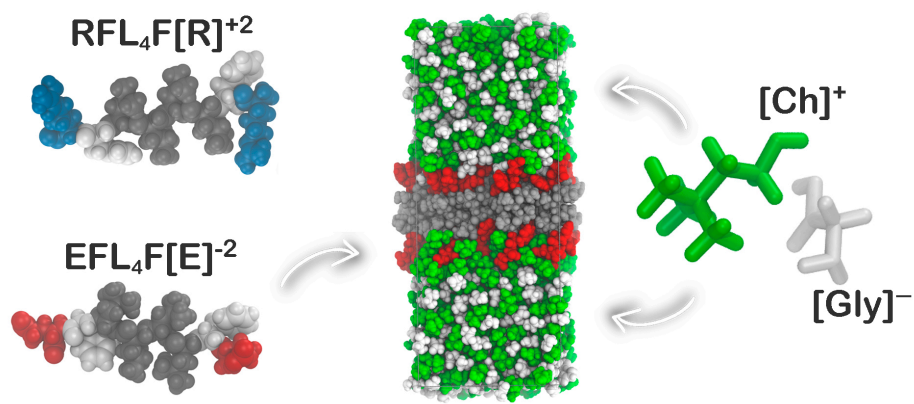

charge. At left, the ionic pair that constitutes the supercapacitor electrolyte. At the middle, the computational cell showing the stabilized structure of the nanosheet (in this case specifically the EFL$_4$FE) in the ionic liquid [Ch][Gly]. The hydrophobic core of the nanosheet is represented in gray, the charged groups in red and the ionic pairs in white and green. It is possible to observe a slight permeation of the cations on the negatively charged surface of the nanosheet. Analogous representation can be obtained for RFL$_4$FR system. More details about XFL$_4$FX nanosheets in [Ch][Gly] ionic liquid are shown in the supporting information material.

Considering the high charge densities on the EFL₄FE and RFL₄FR nanosheet surfaces, in this letter we propose to employ such nanosheets as novel electrode materials in electric double layer capacitors. These electrodes, formed only by amino acids, are naturally completely biodegradable. But, in order to obtain a biologically safe storage energy device, it is desirable that we make use of a non-toxic electrolyte. For this function, we choose an cholinium-based ionic liquid that is readily biodegradable and non-toxic.[20-21] By associating the cholinium cation with an amino acid anion, we further guarantee the low toxicity of the resulting liquid, making it practically harmless. Among the various cholinium-based ionic liquids recently investigated, we chose the one that employs glycine as anion, [Ch][Gly] (see Fig. 1), because it presents the best balance of desirable properties: toxicity, viscosity and ionic conductivity.[20-21] In addition, this liquid is strongly alkaline (pH 10.3 at 5 mM), a desirable feature here, since we wish to stabilize the protonation state of the surface polar charged groups of the nanosheets while maintaining the charge density approximately constant.

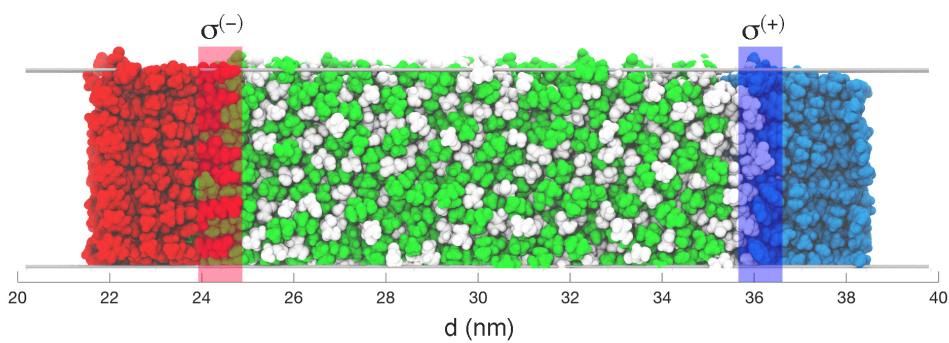

**Figure 2**: Amino acid based supercapacitor. EFL₄FE and RFL₄FR nanosheets as negative (red) and positive (blue) electrodes in red and blue respectively. [Ch][Gly] ionic liquid is presented in white and green. The red and blue stripes indicate negative and positively charged electrode surfaces.

The supercapacitor proposed here consists of an EFL₄FE nanosheet as a positive electrode, and a RFL₄FR nanosheet as a negative electrode as showed at Figure 2. Both nanosheet were prepared to have an approximate area of 17.8 nm$^2$ in ionic liquid which corresponds to a surface charge density of 57.6 μC cm$^{-2}$. This density can be considered high since typical values employed for electrochemical supercapacitor studies are one order of magnitude lower.[22-25] A biophysical system of interest for comparison here are the lipid membranes, for which a charge unbalance at the interface with the aqueous electrolytic solutions is observed, of about 0.01 μC cm$^{-2}$.[26] The separation between charged surface of the nanosheets is about 10.5 nm while the separation between the neutral surfaces (the outer ones, in contact with the vacuum) is 17.1 nm. As can be seen in Figure 2, the charge on the electrodes is distributed in a well-defined plane, so that this device can be modeled as a parallel plate capacitor. Thus, the conventional Poisson equation formalism[27-28] used in previous studies to obtain drop potential, specific capacitance and energy density can be safely used here.

The study of this supercapacitor was carried out through atomistic molecular dynamics simulations. Nanosheets and ionic liquid were modeled using the CHARMM36 based force field.[29] In particular, for ionic liquid we used a Canongia Lopes-Pádua refined model.[30-32] The cells were equilibrated for 10 ns in NPT ensemble to get a consistent mass density at the bulk region. The production stage we have performed a sampling for 50 ns with a 70nm-vacuum slab in the NVT ensemble. Properties were calculated from simulations considering a time-step of 2 fs with coordinates collected every 1.0 ps, which gives a total of 25000 frames for statistical analysis. All molecular dynamics simulations have been performed with the GROMACS 2016 program.[33] Further details can be found in the supporting material.

The stability of both the EFL$_4$FE and RFL$_4$FR nanosheets in the ionic liquid [Ch][Gly] was confirmed by preliminary simulations and the results are presented in the supporting material. In short, it has been found that the polypeptides EFL$_4$FE and RFL$_4$FR interact in one direction by their side chains via hydrophobic contacts whereas in the orthogonal direction they interact by a mesh of hydrogen bonds. In addition, the surface polar groups exposed to the ionic liquid interact by intense electrostatic forces. This inter-peptide network bonds associated with the interaction with the liquid are the factors that confer stability to the nanosheets. The stability observed in [Ch][Gly] is strictly analogous to that occurring in the aqueous environment,[19] except that on the surface of the ionic liquid there is a strong ionic interaction that leads to the formation of a double electric layer. It is precisely this feature that makes these nanosheets attractive and viable for use in energy storage devices.

The malleability of the highly charged surface of the nanosheets gives the supercapacitor a mass distribution significantly different from those observed in planar supercapacitors based on flat electrodes such as graphene, graphite or gold. Figure 3 shows the mass density profiles for the charged nanosheets in contact with the ionic liquid. Clearly it is possible to observe a higher density of cations at the interface of the negatively charged electrode (in red) while the opposite is observed for the positively charged electrode (in blue). However, the most interesting thing to note here is that both electrodes allow a semi-permeation of the ions on their surface, characterized by the overlap between the density profiles of the nanosheets and the ionic liquid. This semi-permeation is similar to that occurring in porous electrodes, where a much larger surface area is available for electrode/electrolyte interaction, greatly favoring the storage properties of the supercapacitor.[34-36]

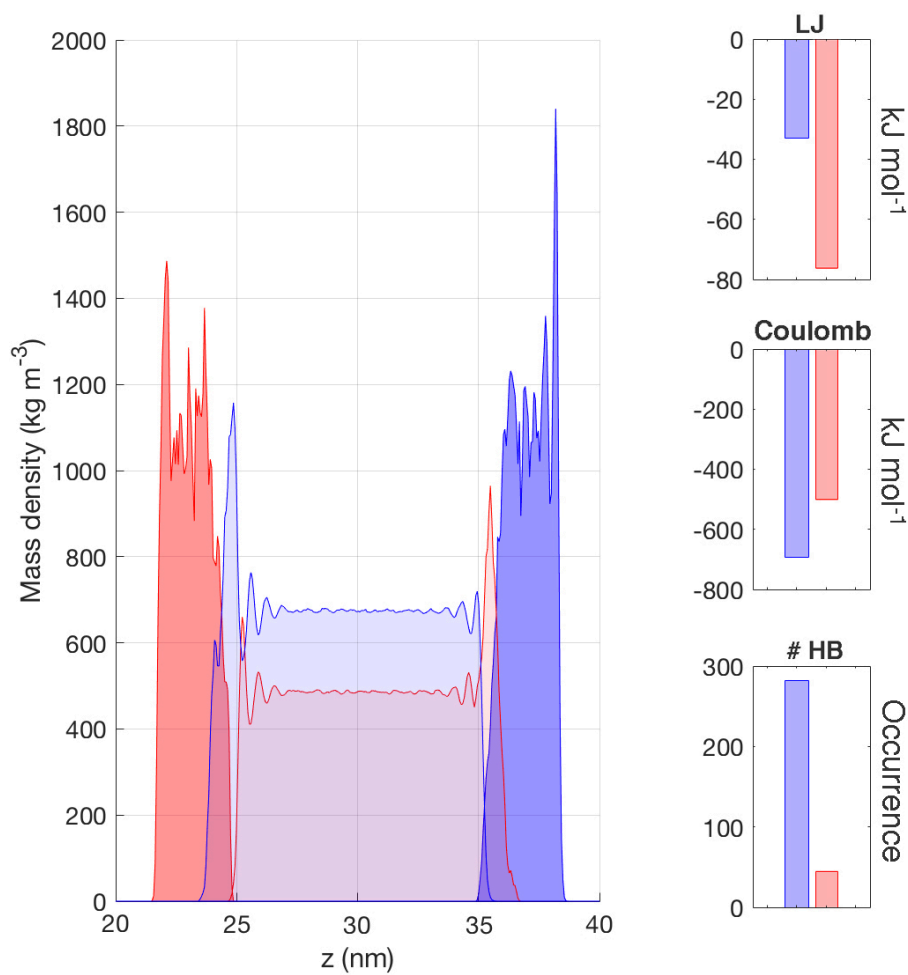

**Figure 3:** Mass density profiles (kg m⁻³). The mass distributions related to the nanosheets are in darker colors (negative electrode in red and positive electrode in blue). The lighter curves refer to the ion distributions, where light blue and light red refer to the cation [Ch]⁺ and the anion [Gly]⁻ respectively. At right are presented van der Waals and Coulomb components of the interaction energy (kJ mol⁻¹) and the hydrogen bond occurrence at the electrode-electrolyte interface. In the plots, the bars refer to the nanosheet-ion interactions; being the blue bar associated with the [RFL₄FR]/[Gly] interactions while the red bars to the [EFL₄FE]/[Ch] interactions.

The intense interaction between electrode and electrolyte can be described in energetic and structural terms through the analysis of the formation of hydrogen bonds with the interfaces. Such analysis is done through the correlation between the pairwise interaction energy and the hydrogen bond occurrence in the interface electrode-electrolyte. It is not a thermodynamically rigorous treatment, since the pairwise energies do not consider the entropic contributions for interaction, taking into account only the enthalpic part. But it is possible to obtain important information about the behavior of the electric double layer, which is strongly ruled by interactions between the nanosheet charged surface and their corresponding counter-ions. In the bar plots of Figure 3 we observed that, as expected, the electrostatic contribution is much more relevant for the formation of the electric double layer than the van der Waals contribution. Its magnitude is approximately an order of magnitude superior, easily observed by comparing the vertical scales of plots. The blue bars refer to the interaction between the glycine anion and the [RFL4FR] positively charged electrode; [RFL4FR]/[Gly]. This interaction, of -694 kJ mol$^{-1}$ (per polypeptide) is significantly larger than its corresponding, [EFL4FE]/[Ch], of -492 kJ mol$^{-1}$. These values are related to the number of charged sites in each peptide-ion interagent system and are also consistent with the number of hydrogen bonds formed at each interface. The total number of electrode/counter-ion hydrogen bonds at each interface was 282 and 45 for [RFL4FR]/[Gly] and [EFL4FE]/[Ch], respectively. The strong interaction at the interface associated with the large number of hydrogen bonds formed between electrode and electrolyte are remarkable features that may exploited to increase or adjust the capacitance of the device. The first aspect to be observed is the large difference between the interactions that occur at both the positive and negative interfaces. As we will see later this difference has a strong impact on the properties of the supercapacitor, conferring it a great asymmetry and therefore directly affecting its capacitance. Another aspect concerns the effects of pseudo-capacitance. Such effects are related to faradaic reactions groups containing oxygen or nitrogen and have been reported as an increase factor of the total capacitance.[34, 37] Possibly these effects could be obtained for the polypeptide electrode supercapacitors through variations of pH, for example.

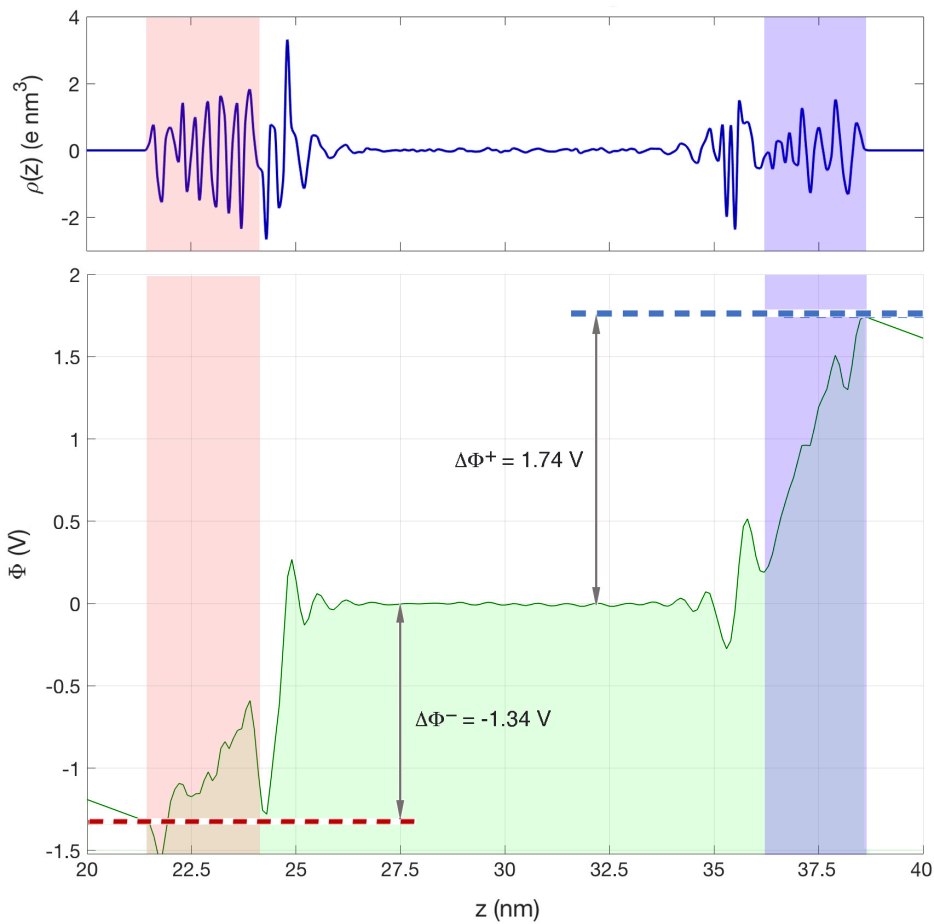

**Figure 4**: At top, charge density distribution ($e$ nm$^{-3}$). At bottom, the electrostatic potential (V) through the supercapacitor. The dashed lines indicate the reference for the calculation of the electrode potentials ($\Delta\Phi^+$ and $\Delta\Phi^-$), in relation to the ionic liquid bulk in the center of the supercapacitor. The red and blue stripes indicate the position and thickness of each electrode, negative and positive, respectively.

The excess charge on the polypeptide electrodes induces a redistribution of charges near its surface. This spatial charge density, which is directly related to the potential through the supercapacitor, is given at the top of Figure 4. We can observe that the charge density distribution, $\rho(z)$, suffers severe oscillations inside the electrodes (within the blue and red stripes), which does not occur for typical planar electrodes (such as graphene, graphite, gold, etc.).[22-25, 38-39] These oscillations originate from the hydrophilic carbonyl oxygen and amino hydrogen sites distributed along the polypeptide chain, which are responsible for the electrostatic inter-peptide interactions in the nanosheet. It is interesting to note that the potential drop inside the electrodes contributes to the total potential drop across the supercapacitor.

The electrostatic potential, $\Phi(z)$, through the supercapacitor can be obtained by numerical integration of the one-dimensional Poisson equation:[27-28]

$$\Phi(z) = -\frac{\sigma z}{\varepsilon_0} - \frac{1}{\varepsilon_0} \int_0^z (z - z') \, \rho_z(z') dz'$$

Figure 4 presents the electrostatic potential profile. This potential decrease linearly outside the limits of the supercapacitor and becomes constant it its interior near the center region, around 30 nm. In the vicinity of the electrodes it presents characteristic peaks related to the formation of the electric double layer. With the exception of the potential difference observed inside the electrodes, the general behavior of the potential across supercapacitor is compatible with that observed for typical graphene-based supercapacitors. The potential difference in electrodes (half-cell potential) was determined as the difference between potentials at the electrode and in the bulk region:

$$\Delta\Phi^+ = \Phi^+ - \Phi^{bulk}$$

$$\Delta\Phi^- = \Phi^- - \Phi^{bulk}$$

Thus, the potential difference across the supercapacitor is given by:

$$\Delta\Delta\Phi = \Delta\Phi^+ - \Delta\Phi^-$$

From Figure 4, we obtain $\Delta\Phi^+ = 1.8$V; $\Delta\Phi^- = -1.3$V and $\Delta\Delta\Phi = 3.1$V. Here it is important to note that despite the large surface charge density stabilized on the polypeptide electrode surfaces, the potential difference across the supercapacitor presents a typical value, lying within the electrochemical window of the conventional ionic liquids, ie, less than 6V.[40]

The specific capacitance of the electrodes, $C^+$ and $C^-$, depends on the surface charge density at the electrode ($\sigma$) and the potential drop from electrode to bulk. As mentioned before, the surface charge density is taken as the average total charge on the electrode divided by its surface area (resulting in $57.6\ \mu C\ cm^{-2}$). Thus, the integral specific capacitance can be calculated as: $C = \sigma/\Phi$.[27-28] From this relation, we obtain that the electrode specific capacitances are $C^+ = 33.1\ \mu F\ cm^{-2}$ and $C^- = 43.0\ \mu F\ cm^{-2}$. The total specific capacitance through the electrode is obtained by: $1/C_T = 1/C^+ + 1/C^-$ which results in $C_T = 18.7\ \mu F\ cm^{-2}$.

The observed difference between $C^+$ and $C^-$ reflects the inherently asymmetric nature of this supercapacitor. This asymmetry originates from two different sources whose effects are strongly coupled and inseparable: the considerable structural difference of ions [Ch]$^+$ and [Gly]$^-$ and the chemical composition of the electrodes consisting of two different polypeptides. As we saw before, the anion-positive electrode interaction is significantly higher (about 40%) than the cation-negative electrode interaction. The higher interaction of the ionic liquid with the positive electrode promotes a higher concentration of charges on its surface, resulting in a greater potential difference and, therefore, a lower electrode capacitance. This explains the unusual fact that $C^-$ is larger than $C^+$, although the opposite is usually observed. A large volume of previous results for flat electrodes/ionic liquid supercapacitors has been found a larger value for the positive electrode capacitance rather than the negative one. [22-24, 38, 41]

The specific capacitances obtained for this polypeptide supercapacitor are relatively high. For example, the electrode capacitance is considerably higher than those reported for graphene-based supercapacitors with conventional ionic liquid as electrolyte. Kim et al. investigated a graphene supercapacitor using a mixture of [EMIM][BF4]/acetonitrile as electrolyte and found for $C^+$ and $C^-$ the values of 3.62 $\mu F\ cm^{-2}$ and 3.28 $\mu F\ cm^{-2}$ at a potential difference similar to that employed here, $2V$.[25] For total capacitance the value found is also higher. The authors found a value of 2.6 $\mu F\ cm^{-2}$ for the total capacitance using pure [EMIM][BF4] as electrolyte.[23-25, 42] Slightly smaller values (in the range 1.9-2.3) were obtained when graphene was replaced by graphene oxide.[23] Similar results, for a large diversity of graphene-based/ionic liquids supercapacitors, were found for the total capacitance, all presented values ranging from 3 to 8 $\mu F\ cm^{-2}$.[23, 28, 36, 38, 43-51]

Higher values for total specific capacitance were obtained only when confinement effects were present, as in the case of electrolytes in slit or pores[52-56] or when morphological variations, such as edge effects and roughness on the electrode surface, were considered.[55, 57] The results found here are consistent with the latter case. That is, the polypeptide electrodes surfaces are malleable, rough and allow a semi permeation of the counter-ions, dramatically favoring the access of the ions to the charged sites of the electrode which implies in increasing the capacitance of the device.

In order to examine the influence of the polypeptide electrode on the supercapacitor properties, we performed a simulation for a planar gold electrode supercapacitor using the same electrolyte, [Ch][Gly]. For a physically justified comparison we use the same total potential difference founded for the polypeptide supercapacitor; 3.1V. The details of this simulation are given in the support material and the results of interest are presented in Figure 5.

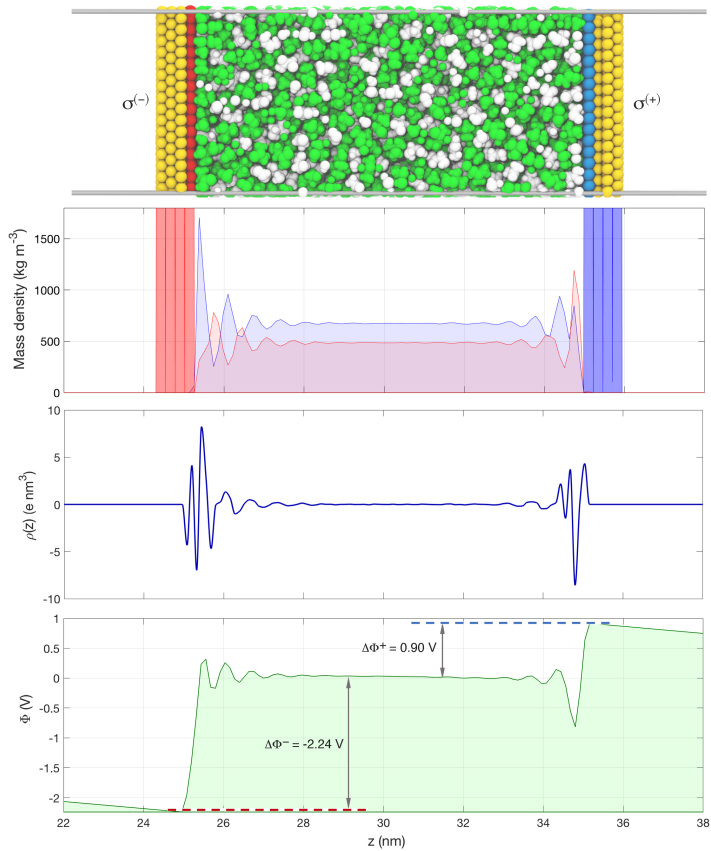

**Figure 5**: Planar supercapacitor with gold electrodes. At top, a molecular representation of the supercapacitor with the electrolyte [Ch][Gly] showed in green and white and the electrodes in yellow. The total charge of the electrode was distributed in the atomic layer in contact with the liquid. Mass density (kg m$^{-3}$), charge density distribution ($e$ nm$^{-3}$) and the electrostatic potential profile (V) are showed in this order from top to bottom.

The analysis of the mass density profiles, charge density distribution and electrostatic potential reveals that the flat gold electrodes supercapacitor presents typical behavior and similar properties to the supercapacitors reported in previous studies.[12, 22, 58-59] For instance, the oscillatory and out-of-phase behavior of the ion distribution in the electric double layer can be observed in mass density profiles. Furthermore, we see that the charge density distributions near to the electrode surfaces present higher probabilities for the ions with opposite charge of the electrode and also that the potential at the negative electrode has a higher magnitude than that of the positive electrode, leading, as expected, to a higher capacitance of positive electrode. For this Au-[Ch][Gly]-Au supercapacitor the values for the capacitances $C^-$, $C^+$ e $C_T$ were 9.3, 3.8 e 2.7 μ$F$ $cm^{-2}$, respectively. This value for $C_T$ is 7 times lower than that obtained for the polypeptide electrode supercapacitor.

In this letter, we use self-organized charged polypeptide nanosheets as electrodes to an electric double layer supercapacitor. This polypeptide supercapacitor, presenting some distinct characteristics such as the semi-permeation of ions inside the electrode surface and a greater capacitance of the negative electrode in relation to the positive. But more important, it presents a total capacitance that is 7 times greater than its analogue using planar metallic electrodes. In terms of energy density ($u = C(\Delta\Delta\Phi)^2/2V$, where V is the volume of the supercapacitor) the difference in performance of the two supercapacitors is even greater. The energy density stored in the polypeptide supercapacitor is 16.7 pJ cm$^{-3}$ while the gold-based one presents as density of only 0.9 pJ cm$^{-3}$. This impressive result is certainly a motivating factor for future experimental and computational investigations. Among the important aspects which have yet to be elucidated are, for example, the reduction of the alkalinity of the ionic liquid, desirable to render the device fully biocompatible. However, the variation of the pH by means of the dilution of the ionic liquid can lead to the loss of efficiency of the supercapacitor. Our preliminary tests, with small molar fractions of water in the electrolyte, indicate that the effect of pure water on the electrolyte is drastic with water permeating inside the nanosheet electrodes, leading to the loss of the double electric layer structure and a great reduction in the capacitance (see supporting information). Another important aspect it is the investigation of the performance of the supercapacitor considering the asymmetry of the supercapacitor. As the lower capacitance ($C^+$) dominates the total capacitance ($C_T$), therefore the development of strategies to increase $C^+$ is crucial to increase even more the capacitance of the supercapacitor.

The device proposed here presents a high potential for energy storage using only biodegradable organic matter. It consists of a prototype for a new class of supercapacitors that can serve as a basis for important applications in several areas, particularly in the development of devices for bio-integrated electronics.

**Information**

Further technical aspects and structural analysis are also provided. This material is available free of charge via the Internet at http://pubs.acs.org.

**Acknowledgments:**

This work was supported by grants from Brazilian agencies FAPESP (Grant number: 2017/11631-2), FAPEG (Grant number: 201610267001030) and CNPq. E.E.F is particularly grateful for the contribution from Prof. Marcos Quiles.